# Enabling Responsible, Secure and Sustainable Healthcare AI - A Strategic Framework for Clinical and Operational Impact

Jimmy Joseph *

*Solutions Engineer Advisor Sr., United states.*
jimsweb@gmail.com

**Abstract**

Healthcare AI will realize long-lasting clinical and operational improvements only if technical innovation can be combined with organizational strategy, governance, security, and continued adoption. We describe a five-pillar strategic model - Leadership & Strategy, MLOps & Technical Infrastructure, Governance & Ethics, Education & Workforce Development, and Change Management & Adoption - for responsible AI in healthcare. Unlike other guidance, our approach includes AI-specific security and lifecycle MLOps combined with cross-disciplinary governance and user-centered change management - aligning a "compliance-by-design" philosophy with emerging regulatory expectations. We demonstrate the framework in two practical applications: (A) a hospital service for inpatient length-of-stay prediction and (B) an AI-assisted radiology second-reader for lung nodules. The LOS model achieved $R^2 \approx 0.4$–$0.6$ in pilot evaluation across cohorts and was used by >75% of case managers, with targeted units implementing 5–10% reductions in average LOS for complex-discharge patients. A radiology tool (sensitivity ≈ 95%) was embedded in PACS with thresholding and explanation overlays to decrease alert fatigue, resulting in an 8% increase in detection of sub-centimeter actionable findings and no significant read slowdown. Overall, AI services were run under monitored and auditable pipelines in both cases, with no reported security incidents, and user-facing model cards informed trust and appropriate use. These findings show that robust MLOps and security, when coupled with governance, education, and change management, significantly enhance both acceptance and impact of AI at the bedside as well as in operations. We conclude by considering limits, generalization, and a roadmap for scaling. LOS pilot (n=3,184 encounters across 4 adult units at a single U.S. hospital, June–August 2025) achieved $R^2$ = 0.41–0.58; case-manager adoption reached 78% by week 6; targeted units observed 5–10% relative reductions in mean LOS for complex discharges versus each unit's pre-pilot baseline. The radiology second-reader pilot (n=1,126 chest CTs over 8 weeks) showed +8.0 percentage-points in sub-centimeter actionable nodule detection (95% CI [2.1, 13.9]; $\chi^2$ p=0.008), with median report turnaround time unchanged at 23 min (Wilcoxon p=0.64).

## 1. Introduction

Healthcare, in particular, is seeing a surge of interest and investment in AI/ML technologies, driven by the emergence over the past few years of powerful models (e.g., generative AI) and a profusion of clinical data. This wave of innovation has led stakeholders at all levels (from day-to-day practitioners on the front lines to C-suite executives) to ask what role they can play in realizing AI within practice and how to balance them against their risks[1]. Early wins in AI, such as medical image analysis and predictive analytics, have shown huge potential for advancing patient outcomes and operational effectiveness. Deep learning models, for instance, now achieve near-human performance in diagnosing diseases from high-resolution medical images (e.g., screening diagnostics such as diabetic retinopathy in eye examinations or tumors in radiological analysis) [2][3]. This sort of analysis can be performed by ML models, which can predict hospital-relevant metrics such as patient length-of-stay, assisting administrators in capacity planning and resource allocation[4]. These progressions are suggestive of the transformative impact of AI in diagnostics, treatment decision support, and healthcare management.

But the promise also comes with well-documented challenges that have prevented AI tools from moving from pilots to clinical workflow. There are a number of technical and organizational reasons that many healthcare AI projects fail to have enduring impact[26]. Typical fail points are misconceptions of use case, lack of leadership buy-in, lack of competence, bad data quality, and poorly integrated workflows[6]. Moral and ethical concerns like algorithmic bias, black-box model opacity, and patient safety harm have also raised concerns. AI systems that are opaque or make surprising decisions can discourage patients and clinicians from embracing them and limit adoption. Additionally, AI brings new security threats - for example, adversarial attacks might tamper with a model's inputs and outputs with potentially harmful repercussions in a clinical environment[23]. These complex problems illustrate that the successful adoption of healthcare AI is not only a data science challenge, but also a socio-technical initiative with strategic considerations.

In this paper, we present a strategic approach so that AI can be implemented in healthcare in a responsible, secure, and sustainable manner. This framework builds upon current literature and experience - in particular, Monga's writings around responsibly, safely, and sustainably navigating the "AI/ML maze" within healthcare[7], as well as Singh et al. on deep learning for biomedical signal and imaging, to draw axes of success. We define five foundational components of implementation: (1) Leadership & Strategy, (2) MLOps & Infrastructure, (3) Governance & Ethics, (4) Education & Workforce Development, and (5) Change Management & Adoption. In each, we detail leading practices for AI initiatives that deliver clinical and operational impact, maintain ethical considerations, and mitigate concerns.

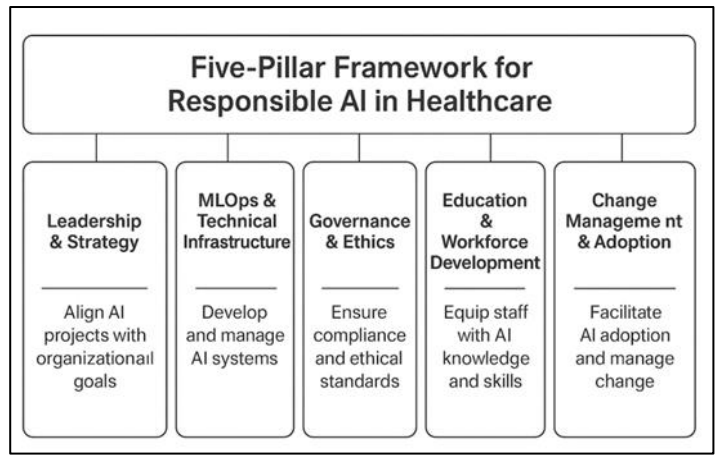

The rest of the paper is organized as follows. Background: We present the background to the state of play in AI adoption for healthcare and why there is a clear need for ethical, explainable, and trustworthy AI today, as well as regulatory trends. Methods: We describe the proposed framework and its components. Results: The Results section discusses real-world deployment considerations and how the framework addresses these challenges through two use cases: a hospital length-of-stay prediction model and an AI tool in radiology for decision support. Discussion: We give a wider discussion of implications, including the merging of biomedical signal and imaging AI, future trends (e.g., generative AI), and lessons learned in scaling AI clinically. We end with suggestions and future perspectives for enabling impactful and trustworthy AI at large scale in healthcare.

## 2. Background

### 2.1. AI's Growing Role and Challenges in Healthcare

AI and ML have evolved significantly in recent years, with solutions for challenging healthcare issues increasingly developed. In the medical field, deep learning algorithms can now outperform humans in some diagnostic tasks. Indeed, most AI-based medical devices that have been granted regulatory approval to date are based on imaging data; for example, more than 84% of all AI-based medical devices cleared by the US FDA employ medical images, primarily from radiology, with a smaller proportion working on physiological signals and electronic health record data[8]. These trends mirror the strong alignment of deep learning (DL) models for image analysis and enthusiasm in the radiology community for AI tools, exemplified by detection of diabetic retinopathy on fundus photographs and lung nodules on CT scans, as tools that can be leveraged to enhance clinical decision-making[2][9]. Beyond imaging, AI tools applied to biomedical signals (such as ECG or EEG waveforms) are beginning to evolve for early detection of cardiac arrhythmias and seizures among other conditions - often achieving performance that is on par with expert interpretation[10]. Elsewhere, predictive models applied to hospital administrative and clinical data have demonstrated potential for predicting outcomes such as length-of-stay and risk of readmission[53], which in turn can be used to anticipate the need for intervention and resource allocation[5]. These progressions are suggestive of the transformative impact of AI in diagnostics, treatment decision support, and healthcare management.

However, despite these advances, there remain a number of challenges that must be overcome to make AI have a real-world impact in healthcare. Quality and bias are consistent challenges - models trained on retrospective or institution-based data may fail to work well in new environments, introducing bias that can worsen disparities in health. Complex model types (such as deep neural networks) suffer from a lack of interpretability, also known as the "black box" problem[10]; it is not readily apparent to stakeholders how an AI arrived at its recommendation. This opacity erodes trust and accountability. Healthcare is an area where decisions can have life-or-death consequences, and so clinicians and regulators require at least some degree of explainability - in fact, the lack of it has been a basis to oppose some AI algorithms for clinical use[10]. Indeed, in some instances in patient care, a less accurate but interpretable model is preferred to a difficult-to-understand and complex tool[14]. This has led to a growing interest in explainable AI (XAI) methods for health use cases, but adopting these approaches into practice is difficult.

Ethical and governance challenges represent another significant barrier. Bias in training data can be harmful, resulting in unfair and even unsafe outcomes for minority groups who are underrepresented in models. In the absence of intentional regulation, AI might be used in ways that compromise patient privacy or informed consent. Global organizations and panels of experts (e.g., OECD, WHO)[28] have set forth principles for responsible AI in healthcare - such as transparency, accountability, fairness, human oversight, data privacy, and safety - as guiding principles to the development and adoption of AI technology[16]. Venturing from high-level principles to operationalization requires specific organizational processes and oversight mechanisms, which many health organizations still lack.

Moreover, security threats specific to AI/ML systems have been revealed[45]. One can think of "adversarial attacks" as malicious inputs that cause a model to give the wrong outputs: in medical images, for example, this might involve very slight changes to an image that bamboozle a diagnostic algorithm. Additional attack vectors include data poisoning (tainting the training data to influence the model), model inversion (recovering sensitive training samples from a model)[13], and even outright model theft. Healthcare AI is a prime target for cyberattacks, with increasing reports of security breaches targeted at AI[18]. The potential consequences of a compromised clinical AI system are significant: incorrect diagnoses, inappropriate treatments, privacy breaches, and regulatory failure. Classical IT security measures are necessary but not sufficient for AI; in practice, proactive risk estimation and defense methods that can be adapted to ML (e.g., adversarial robustness testing, model-drift or anomaly detection) will be required[19][20].

These technical problems are exacerbated by organizational and human considerations. Most AI pilot projects never become lasting clinical programs because of lack of workflow integration and change-management issues. Frontline staff may resent a tool that they consider cumbersome, incomprehensible, or an invasion of their professional judgment. Technology, no matter how good the algorithm, will remain unused if training and clinical leadership are not engaged. What's more, healthcare AI applications tend to involve team-based efforts across functions (IT, data science, clinicians, compliance), and if there isn't harmony among these groups, integration can lose traction. Healthcare leaders surveyed have cited a lack of stakeholder buy-in, inadequate readiness and skills training deficits, and ineffective communication as top reasons AI initiatives fail to achieve expected ROI[6]. The combination of social and technical factors is estimated to reduce the chances for a significant return on investment by 40–50%[21].

Finally, the question of sustainability in AI solutions is emerging. In this context, sustainability means being able to keep AI systems running and scale them over time so they continue to provide value - not just deploy once and let them sit fallow or rust. Model performance can change over time as clinical practice or the patient population evolves, thus requiring ongoing monitoring, updating, and re-validation of models (often referred to as "MLOps" practices). The AI environmental sustainability agenda, including the computational energy requirements to train models, has been noted[50]. But in healthcare there is particular interest in sustainability because of its focus on maintaining operational value. An AI solution that is not deeply integrated into the organization with clear ownership and momentum behind it will be unlikely to survive long beyond the initial wave of excitement. Thus, planning for maintenance over time and consideration of system evolution should be addressed from the start. It is also important to align AI initiatives with business goals.

In conclusion, the proposed benefits of AI in healthcare are vast, but realizing these promises requires resolving issues related to ethics, security, explainability, and adoption. A systemic approach is required that considers both technical strength and human context. Recent works have started to map pathways for this. Monga (2025), in particular, highlights a triptych of must-dos - building AI systems responsibly (with ethical guardrails), securely (with defenses against threats), and sustainably (with processes for continued value creation) - as foundations to success. Singh et al. (2025) have shown that high-performance AI models must integrate domain knowledge and rigorous validation in order to be useful for medical practice, using multiple biomedical imaging case studies. Leveraging these insights, we present an integrated framework that instantiates these concepts into practical strategies for healthcare organizations.

**2.2. Regulatory and Policy Landscape**

It should be mentioned that the regulatory landscape more generally, including in contexts such as healthcare where real-world impacts are paramount, is moving towards mandating responsible AI practices. Within the United States, the FDA has ramped up scrutiny over AI/ML-powered medical software by promoting practices such as algorithm transparency, post-market surveillance, and bias abatement for AI as a medical device. In the European Union, most healthcare AI systems (notably diagnosis or treatment) will be deemed "high-risk" under the new EU AI Act[31] and will be placed under stringent requirements concerning risk controls, training data quality, transparency regarding training data, human oversight, and continuous monitoring[5]. For instance, EU businesses that offer high-risk AI under the EU AI Act are required to build in risk-management systems, require high-quality, non-discriminatory datasets, maintain exhaustive technical documentation, and ensure human oversight or control in an AI's operations[22]. What is clear from these regulations is that robust governance, documentation, and human supervision - as incorporated into our framework - are not only good practice but will soon also be legal requirements.

Furthermore, international standards bodies and professional associations are also promulgating guidelines and frameworks (for example ISO standards on AI or the World Health Organization's guidance on AI ethics and governance in health). Together, these trends suggest a future that will require responsible AI development and deployment. Healthcare organizations that build expertise early on in responsible AI - with cross-disciplinary governance councils, compliance processes, and adherence to frameworks such as the one we propose here - will be far better equipped to confront regulatory requirements and sidestep implementation pitfalls. With this background established, we now describe our approach for enabling AI in healthcare, placed within that strategic framework and echoing these drivers of success.

## 3. Methods: Conceptual Framework for Responsible AI Implementation

To tackle the difficulties detailed in the previous section, we have designed a high-level structure which consists of five vertically integrated cornerstones: Leadership and Strategy, MLOps and Technical Infrastructure, Governance and Ethics, Education and Workforce Development, and Change Management and Adoption. Each pillar corresponds to a key dimension of capability that health systems will need to build in order to deploy AI at scale effectively.

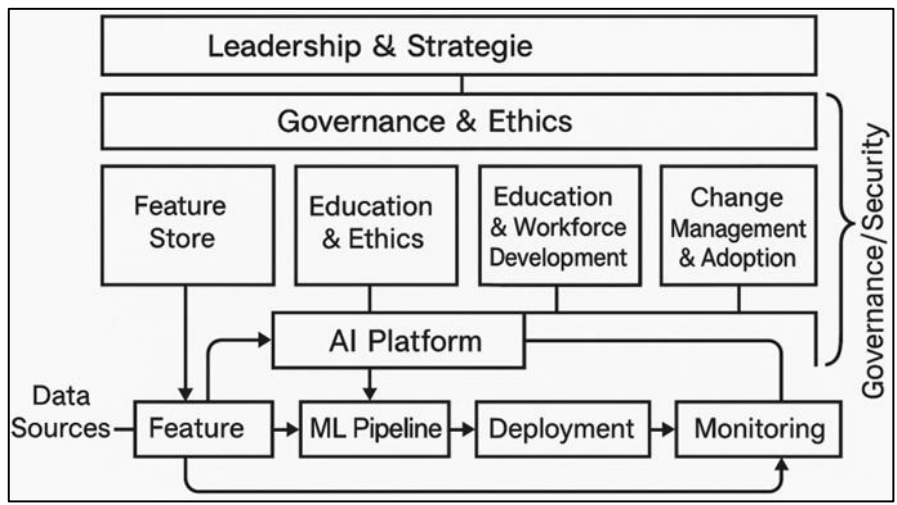

**Figure 1** Responsible AI framework

Figure 1 shows the architecture of the framework and how these components interact with each other. In what follows, we describe each pillar and present the essential practices and structures associated with it, through integration of insights from practical evidence in healthcare AI case studies and expert recommendations.

### 3.1. Leadership and Strategy

Powerful leadership is the foundation of any significant healthcare innovation at scale, and AI is no different. Leadership and Strategy in healthcare AI is when executives and clinical leaders lead their organization on AI initiatives, ensuring that these are aligned with organizational objectives, enabling objective-setting guidance, and creating a top-down culture of support. It starts with having a strong AI vision or charter articulated at the C-level. Health systems should establish an AI/ML steering committee or center of excellence (CoE) with representation from C-suite leadership (e.g., a Chief AI Officer or Chief Data Scientist, in addition to the CIO, CMIO, CMO, etc.). These leaders are also in charge of clearly communicating how AI enhances the delivery system's mission (e.g., better quality, efficiency and patient experience) and leading the drive for resources (finance, people and technology) to achieve that future state.

A key leadership function is instilling focus around AI use cases that support strategic goals and have a clear path to implementation. Instead of pursuing every shiny AI application, top firms "start with the end in mind" by first choosing projects that have a well-defined value proposition and predetermining how success will be measured[25]. Leaders need to support cross-discipline assessment of potential AI use cases (e.g., clinical relevance, data readiness, workflow fit and ROI) before initiating development. This well-defined intake and prioritization process is meant to avoid technically interesting, strategically misaligned projects that aren't practically viable.

Leadership should also work to weave moral principles and risk consciousness into the AI strategy. Many organizations are developing their own set of AI ethics or "AI use" principles, which usually refer to international guidelines, to guide all of their AI projects, such as a "north star"[16]. For instance, an institution may decide to use only AI models for which there is a sufficient degree of explainability from the perspective of end-users (thereby allowing clinicians to interpret model outputs)[17]. Fairness, accountability, and human oversight are also among the principles that must be formalized and publicly articulated. Leadership can and should leave it to the AI governance body (addressed next) to operationalize these principles over time and keep tabs on compliance. This top-level commitment from leadership demonstrates to all staff that the pursuit of AI will be conducted in a considered, responsible way, not a technology-push-at-any-cost approach.

A second area of leadership activity is to nurture an AI-ready culture. This implies the promotion of a data-driven culture and innovation across an organization. Leaders should promote early "wins" of AI projects to create momentum, and at the same time be open about failures and learnings, making learning part of normal working life. Because AI implementation may involve changes to existing, well-entrenched processes, it will probably require leadership with vision who can galvanize the consensus required to overcome inertia or resistance. That includes explicitly stating how AI will complement, not replace, people - for example, by explaining that AI can duplicate predictable analytic work so that clinicians can focus on spending more time with patients.

Significantly, leadership needs to invest in the organizational building blocks for AI enablement. Several health systems have seen benefit from the creation of a dedicated AI-focused leader (such as a Chief AI Officer)[27,28] or formalized an AI Centre of Excellence to consolidate expertise and resources. The point is that someone needs to own and be accountable for setting and driving the AI agenda. This management team oversees cross-functional activities, manages AI projects, and acts as a liaison between technical teams and clinical operations. We emphasize that leadership guidance is not a one-off like executive sponsorship: it requires continuous executive support throughout the lifetime of AI initiatives to overcome obstacles, including addressing inter-departmental tensions and authorizing policy changes (e.g., on data sharing or IT investments).

In conclusion, strong leadership is better able to guide AI direction and orient it towards goals and values, activating the organization for the adoption of AI. Without such leadership, even the best AI technology is likely to have little system-level effect.

### 3.2. MLOps and Technical Infrastructure

The second pillar, MLOps and Technical Infrastructure, encompasses routines and tools that lead to the reproducible building of AI/ML models in a healthcare setting[15]. "MLOps" (Machine Learning Operations) is a field based on DevOps, which drives operational best practices around machine learning lifecycle needs in order to ensure that models are deployable and updatable in an automated, reliable manner[29]. In an industry like healthcare, where the safety of patients and the privacy of data are a priority, a clearly defined MLOps pipeline is essential to move prototypes into production systems that clinicians can rely on. Ongoing monitoring (Algorithm 1) detects drift and triggers controlled responses.

**Algorithm 1**
```
SCHEDULE: run daily/weekly
INPUT: recent_predictions P_t, outcomes Y_t (when available), reference_baseline B

metrics ← {AUC, MAE/R2, calibration, PSI(data), drift(embedding)}
FOR m in metrics:
    delta[m] ← compare(m(P_t, Y_t), B[m])
```

```
IF any delta[m] > alert_threshold[m]:
    create_incident(ticket, attach_artifacts, notify {DS, Product, ClinOwner})
    IF safety_metric_breach: trigger_safeguard (rollback_or_shadow)
LOG all metrics, deltas, actions for audit; update dashboard
```

A strong MLOps framework will cover the following components: a) version-controlled workflows for model development, which include code repositories and experiment tracking; b) rigorous validation and testing processes, including - but not limited to - validation on separate data sets, bias and performance audits of models, and simulation in a clinical setting; c) automated deployment pipelines that can move models from staging environments to production services (e.g., via containerization or APIs); d) monitoring systems that can continuously track the performance of an ML model (accuracy, data drift, prediction drift, etc.) with triggers if it goes beyond preset thresholds; e) processes for tracking all outputs and relevant data derived from those outputs for later audits (to support explainability and debugging); and f) defined procedures for rollback or updating - such as regular retraining on new data or patching models if errors are uncovered during their lifetime. Integrating these components helps to guarantee that when an AI model is created, it can perform within the clinical IT ecosystem and be sustained over the long term.

Providing suitable infrastructure for these MLOps pipelines should be pursued by healthcare institutions. This might take the form of data management platforms (e.g., a feature store that centrally manages and serves validated features for models[30]), as well as orchestration tools and monitoring for productionized models. As an example, a feature store can cut data-prep duplication significantly while also ensuring that training and inference data are consistent - critically important in healthcare where the definition and quality of data must be under tight control. Setting up a model catalog or registry also permits traceability of all models, their versioning, metadata (such as the training data used, the algorithm type and its hyperparameters), and ownership[32]. Not only is this critical for internal governance, but also for compliance - e.g., knowing at any given point which models are having an impact on patient care.

Security must be built into the MLOps pipeline. Unlike common software, ML models cannot simply "plug into" currently used security scanning tools. Checks for ML-specific vulnerabilities should be added as part of model testing, including whether the model is resistant to small variations in input (adversarial tests)[47] and whether patient data are memorized unintentionally by the model[19]. The literature indicates that many existing industry MLOps pipelines fall short from a security perspective, leaving room for open issues[35]. As such, health organizations will need to extend their dev/test pipelines to encompass AI-specific security evaluation, and partner with cybersecurity teams to modify controls for AI (for example, adding checks that look for abnormal usage patterns of an AI service which could be indicative of an attack).

Another key aspect of MLOps is the ability to continuously learn and improve models. In healthcare, changing populations or practice can lead to reduced deployed model accuracy (i.e., a model whose performance decreases after an important change in clinical protocols or during a pandemic). Our approach emphasizes the importance of ongoing model surveillance and feedback between model outputs and outcomes, allowing new models to be tested as they arise[36]. This can be done with dashboards that contrast model predictions and actuals at various timescales, as well as mechanisms to gather user feedback. For example, if many of an AI's recommendations are overridden or corrected by clinicians, such feedback should be quantified and analyzed by data science staff to help pinpoint possible sources of flaws in a model or the need for retraining.

Infrastructure should be scalable and reliable. In healthcare, for example, we may want models to run in real time or near real time (such as an AI reading a radiology scan in the ER). Architectures should be designed for high availability and low-latency inference - considering the use of cloud services or on-site GPU servers. Given the inevitable security and privacy compliance requirements (since health-related data are very sensitive), such as HIPAA or GDPR, all processing pipelines must be designed from the start to comply[56][36] through proper use of encryption, access policy monitoring mechanisms, audit logging controls, and related safeguards.

In conclusion, MLOps and technical infrastructure are the engines that safely and effectively drive AI models from bench to bedside. Organizations that build these capabilities will be far better positioned to deploy multiple AI applications and scale results. Conversely, the absence of these capabilities is what prevents promising projects from ending up in "pilot purgatory" or unexpectedly flopping in practice due to a lack of support processes.

**3.3. Governance and Ethical Oversight**

The Governance and Ethics pillar guarantees that ethical principles, regulation, and accountability are embedded into AI initiatives. We suggest that healthcare organizations establish an AI governance council or committee to oversee AI/ML initiatives enterprise-wide. This council should be cross-functional, with participation by representatives from clinical leadership, data science/IT, legal/compliance, data governance, security, etc., as is standard in most health systems (e.g., nursing operations)[40][41]. The responsibilities of the governing body are multi-faceted.

Endorsing Use Cases and Tools: The governance committee will assess new potential AI use cases to make sure they align with the organization's strategic objectives and its ethical stance. As mentioned before, leaders bring use case proposals for screening to this committee. The governance group further challenges each submission by asking: Is the clinical/operational need clear? Is the dataset suitable and of high quality? Have biases or risks been taken into account? Is there a clear place where the model output will be used in workflow? Only cases that satisfy certain prerequisite conditions proceed further[42]. This step ensures that experiments without adequate oversight or those that might be harmful will never reach patients. The committee also reviews new AI tools or software platforms for implementation, checking them against regulatory approval, security, and integration with other systems[43][44].

Policy Development: An important responsibility is for AI governance to create the policies, procedures, and best practices around how data will be used in AI development and usage[45][46][48]. These may comprise a policy on AI explainability (any output the AI displays to clinicians must be shown with necessary details, including an explanation or confidence level), policies about how data will be used and consented for model training, validation and documentation of models, and an AI vendor code of conduct. With these ground rules in place, the organization establishes clear lines of responsibility that every AI project follows. Similarly, the governance committee must periodically review and modify these policies as external laws and standards change.

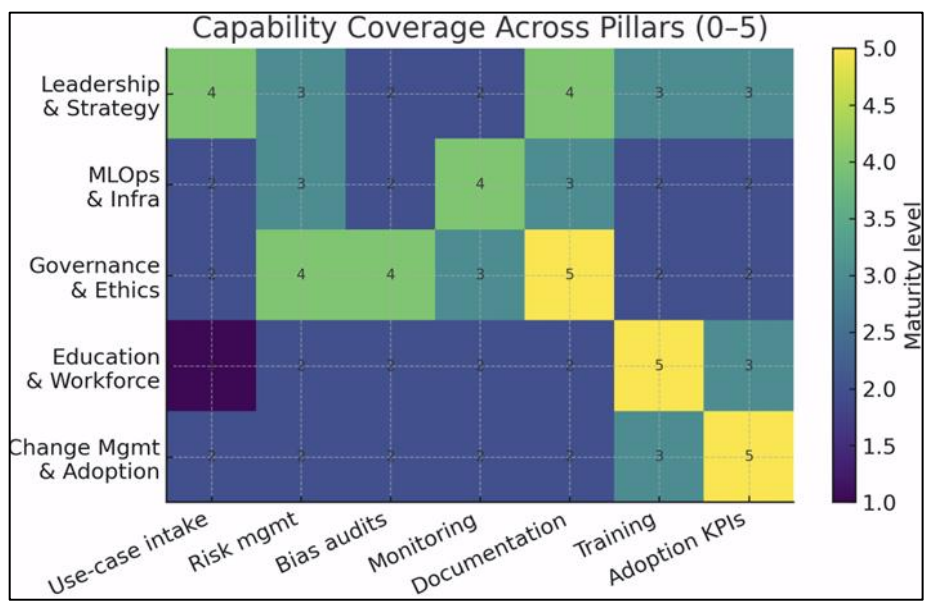

**Figure 2** Government/Pillar heatmap

Continuous Oversight and Risk Monitoring: Governance does not end with the deployment of an AI system; it is ongoing. The committee (or subcommittees it may establish) must constantly oversee adherence to standards and risk management. This may include: managers reading regular reports on how the model is performing and being used; auditing algorithms for bias or drift; and ensuring there is a process for users to report issues they find. For example, if an AI to predict patient deterioration is being utilized, the governance team may call for a quarterly report validating the accuracy of predictions made by the model and any adverse events or near-misses associated with its recommendations. If issues do arise, governance can require mitigation measures or even suspend the AI tool until matters are addressed. In essence, this body provides the human oversight that is one of the hallmarks of trustworthy AI - a way to step in if AI does not behave as it should or when context changes[47]. In the length-of-stay prediction example we discuss, the governance plan consisted of appointment of clinical owners with authority to stop use of AI if abnormal behavior or safety concerns arose[49].

Ethical and Legal Compliance: The governance committee should monitor whether each AI system conforms to healthcare regulations and ethical standards. This means collaborating with the privacy office to ensure HIPAA compliance in all handling of data, assessing whether appropriate patient consent is needed (for instance, for secondary analyses or AI development), and determining whether use of the AI raises any ethical questions - for example, around accountability for automated decision making. If an AI system gives diagnostic or treatment advice, the committee should also examine whether it is a medical device and therefore regulated[58], and whether it needs to be subject to regulatory approval (or notification). Proactive engagement of legal and compliance experts would prevent the rollout of solutions that later face regulatory or trust headwinds.

Documentation and Transparency: Governance must ensure rigor in documentation of all AI models, along with their development methodologies. This documentation should cover, for example: training-data lineage; model algorithm design; validation results; intended use; documented limitations; and appropriate mitigation strategies for relevant risks. This is not only best practice but is increasingly demanded by regulators and standards - for example, the EU AI Act will require technical documentation for high-risk AI systems. Documenting these elements is a forcing function for being explicit about limitations and assumptions. Further, the governing body can mandate that model cards or user-facing documentation be developed for each AI tool (as recommended in responsible AI literature). A model card is a brief overview of what a model is designed to do, the data it was trained on, performance metrics, caveats, and considerations of use for people who access that model[11]. In our use case, the nursing staff were given a user guide (effectively a model card) explaining how the remaining length-of-stay model works and what factors it uses[50]. This approach promotes transparency and fosters user trust.

By institutionalizing such governance structures, healthcare entities establish an internal checkpoint for AI projects to pass that greatly decreases the risks of unanticipated negative consequences. It bakes AI risk management into the fabric of an organization. It is interesting to note that a recommendation appearing increasingly in industry guidelines for responsible AI is to set up an AI governance council. Early adopters report that this helps strengthen interdisciplinary communication and promotes a more thoughtful process of introducing AI[52]. Our model positions governance not as a bureaucratic obstacle, but as an enabler of sustainable AI: it is the vehicle for translating leadership's intent and ethical principles into action - safeguarding durable effectiveness and public trust in AI capabilities.

**3.4. Education and Workforce Development**

The introduction of AI to healthcare workflows changes the roles and skills required of health professionals. Therefore, Education and Workforce Development is one essential foundation for training both technical teams and end-users who will create, understand, and utilize AI tools. There are two sides to this: training the technical talent (data scientists, engineers, and informaticians) on healthcare-specific topics and best practices, and educating clinical and operational staff in AI literacy and usage.

On the technical side, health systems increasingly value talent who possess AI/ML expertise in combination with an understanding of clinical context, regulatory guardrails, and the quirks of healthcare data (like interoperability standards and issues related to data quality). An investment in training data scientists and engineers on issues related to medication, natural language processing, healthcare workflows, or biostatistics may help ensure that AI solutions developed are relevant and safe[54]. Conversely, healthcare professionals with analytics acumen can be taught data-science methodologies - often referred to as "citizen data scientists" if they don't hold official titles of that nature - so they can take part in the development of AI alongside more technical teams or communicate effectively[55]. "In-sourcing" may also be facilitated by sponsored fellowship or rotation programs in organizations where clinicians shadow IT staff and clinicians rotate to the data-science group. If we follow the framework, it makes a case for developing more interdisciplinary learning opportunities to train an AI-and-healthcare-savvy workforce.

AI expertise is essential for the wider clinical and operational workforce. It is not necessary that frontline workers understand the nuts and bolts of how neural networks are trained, but they need to grasp some first principles: what is possible with AI and what is not; how to interpret an AI output; and why issues like bias and validation matter. As part of our framework, we propose sustained educational campaigns about AI. One way to do this is by integrating AI discussions into existing opportunities for medical education - e.g., including elements on AI at grand rounds, CME (continuing medical education) events, nursing education days, and the like. These sessions need to demystify AI, breaking down concepts in lay terms (e.g., what is machine learning, what is an algorithm, what is meant by model accuracy vs. bias)[57]. As Einstein's quote goes, "We can't solve problems by using the same kind of thinking we used when we created them." If we cannot explain it simply, then we do not understand it enough - analogously, the burden is on AI specialists to share fundamental concepts with clinicians in an easy and intuitive manner.

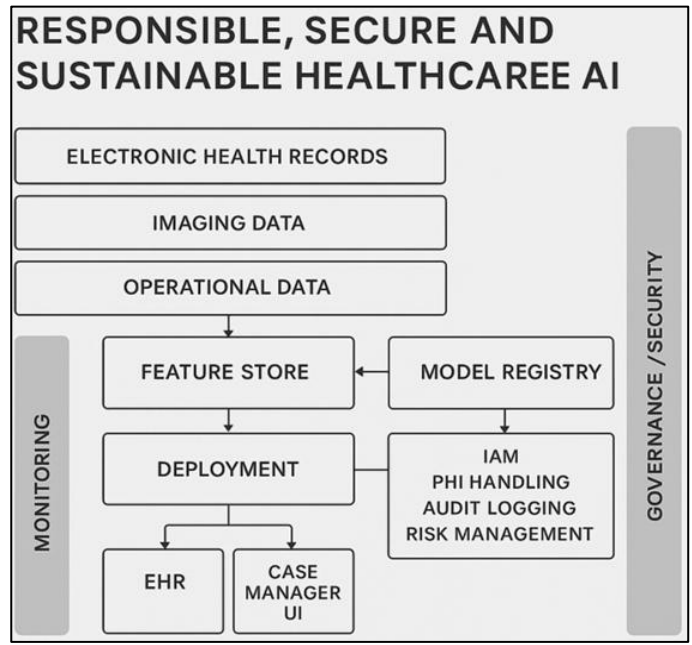

**Figure 3** Responsible, secure and sustainable healthcare AI

In addition to general literacy, training should be available for individual AI tools before they are rolled out. There should be a required official training module or workshop for intended users each time a new AI system is implemented (e.g., an AI sepsis alert in the EHR or a triage tool for radiology images). In doing so, users learn how the AI operates at a high level of abstraction and what its outputs look like, interface features that inform their decision-making process, and-perhaps most critically-they become aware of known limitations of the AI[24]. For instance, radiologists could be trained to know that an AI nodule detector highlights areas of interest on a chest X-ray with a specific sensitivity and specificity, but that they should not rely on it to catch every nodule. In one application, we provided nurse managers and discharge staff with detailed user guides, one-pager desk drops (laminated reference cards), and in-person quarterly introductory sessions plus a gap analysis of typical decision errors versus most impactful clinical decisions[34][51]. These informal sessions gave users the opportunity to ask questions, make suggestions, and learn to use the tool with confidence over time. An iterative educational approach is more powerful than a single training data dump.

We also emphasize the importance of AI champions or super-users being created among employees. Identifying a few people in each department who are super-users, receive more intense training, and act as the go-to for the rest of their departments is an extremely effective way to spread awareness. These champions usually provide support in troubleshooting problems users experience, advocate for the correct use of the tool, and pass user feedback to the AI development team. Peer influence is pivotal to change-management theory - clinicians trust the experience of their peers. Nurse or physician champions who act as evangelists for the AI and help get other people involved accelerate adoption. In the case study, user champions were involved early and remained engaged until completion, which smoothed change management and training efforts.

From a leadership perspective, organizations should embed AI skills into relevant job descriptions and performance expectations. For example, managers in operations may need to be able to interpret data analytics or AI outputs as part of their role. Some health systems have begun measuring baseline AI literacy of staff and are establishing targeted goals for improvement (for example, having all department heads complete an "AI in Healthcare" online course within 12 months). It is also about integrating AI training into on-the-job learning structures as well as leveraging external education resources (universities, online platforms, industry workshops) to reinforce internally established practices. Finally, there is a theme of continuous learning. AI is a rapidly developing field; the technology leading today may be obsolete soon. A robust program should therefore support continued learning. This might include regular AI learning sessions for your workforce (such as an update on "what is new in explainable AI techniques" for the data-science team, or "the potential of GPT-like models in healthcare" for clinical leaders). It is also helpful to encourage employee attendance at suitable conferences or cross-industry forums to keep up with changes. A culture that embraces learning will be more flexible to the changing times ushered in by AI.

In conclusion, the Education pillar of our framework emphasizes the role of the human workforce in keeping abreast of AI technology. By raising AI literacy, creating in-house knowledge, and empowering end-users to learn from and trust AI with appropriate judgment, healthcare organizations can greatly reduce roadblocks to AI adoption, misuse, or misunderstanding of recommendations. An educated workforce is one that is able to work well with AI, using judgment about when to trust the technology and when to challenge it.

## 3.5. Change Management and Adoption

Excellent technology and great governance won't matter if the people who need to use an AI system refuse to use it or circumvent it. Change Management (and Adoption) is the final pillar of our framework, dedicated to discussing how to adopt AI solutions into clinical and operational everyday workflows and how to manage the human side of this change. In healthcare, where workflows are intricate and the stakes are high, implementing a new AI tool often requires strategic planning, communication, and iteration.

A fundamental tenet of change management is getting end-users engaged early and frequently. In the context of AI projects, this means consulting clinicians, nurses, and other front-line staff from the start - not just at go-live. Their feedback is crucial for design (ensuring that the AI output is clinically relevant and displayed accessibly) and for identifying potential barriers to use. For our example length-of-stay application, nursing and case management were consulted during development, and such stakeholders signed off on major decisions around how the prediction would be computed and displayed[63,50]. This early engagement engendered a feeling of ownership and trust: people believed that the tool was created "with them" rather than for them. For every AI project, it is a good idea to create a user-advisory group that can test-drive the tool, provide feedback, and promote it among peers.

Having realistic expectations and verbalizing them is also key. AI shouldn't be sold to users as an infallible or magical solution. Instead, leaders and project managers need to share what AI can and cannot do, and where it sits in current workflows. For example: "This ML model can indicate which patients are likely to need longer stays in the hospital with around 80 percent accuracy. It's designed to help you plan by identifying those patients early, but it's not a replacement for your clinical judgment." By framing AI as decision support rather than decision autonomy, users are more likely to see it as helping rather than threatening. On the implementation side, accessibility is key - consistent touchpoints (live Q&A sessions, emails with tips and tricks, on-call support) will also drive positive uptake.

Workflow integration is a mundane yet decisive consideration for adoption. The AI output must be presented in the correct place and at the time in a workflow when a user can act on it. In many cases, this means harmonizing with established IT systems, such as the Electronic Health Record (EHR) or hospital dashboards, to make AI insights visible in a clinician's typical workflow[49]. If using the AI requires logging into another system or remembering to run a manual query, adoption will languish. In our case the revised length-of-stay prediction was added to the nursing dashboard for patient-flow management already in use, with a clear label that this was an AI-derived value. It was easier for nurses to notice and utilize this prediction, as it was directly in their workflow and required no additional effort. Further, the system offered context including the "factors driving the prediction" next to the score to help make sense of it.

Feedback loops after delivery are essential for ongoing acceptance. Users should have straightforward ways to provide feedback about the AI tool - whether reporting an erroneous recommendation, suggesting refinements, or reflecting on inconveniences. This may take the form of an in-app feedback button, monthly surveys, or meetings with user groups. We advocate creating a cycle where user feedback is examined by the AI team and governance committee, which then alters the AI system or its use-cases in response. In fact, taking user feedback into account is a form of model monitoring: if many users are overriding an AI's recommendations, that could signal that the model needs retraining or that users need further training; governance would want to know what is happening. For instance, the application's development team in the case study reported reasons for not acting on AI scores, which were used to understand and enhance the model. This two-way communication helps members feel heard and creates a partnership perspective.

A third element is handling human reactions and feelings around AI. Some employees may worry that AI will replace their jobs or compromise their professional independence. Openly discussing these fears is critical. Change-management plans need messaging that frames AI as a tool that can automate mundane tasks or offer an added safety net - not as a competitor. At the same time, organizations must be transparent about potential role changes; an AI might, for example, automate some documentation or image-analysis tasks, altering workload. If these transformations are anticipated, HR and leadership can engage in reassigning duties or providing training to turn a potential obstacle into an opportunity for staff to grow into higher-value roles. Monga (2025) recommends working with human resources to develop a workforce-focused change-management plan in the context of potential job-function reshaping with vision AI, including generative models. It is wise to make such arrangements in advance.

Adoption and impact are important for lending weight to change. Establish KPIs for the usage and impact of the AI tool (e.g., percentage of clinicians who use it, reduction in adverse events, or improvements in efficiency). Track these metrics and feedback, and communicate findings to relevant parties. Indeed, when data demonstrates a positive impact - "Since implementing the AI discharge-planning tool, our average length of stay for the target patient cohort has reduced by 0.5 day" - celebrate this and, more importantly, attribute it to both the technology and the people who adapted to it. Acknowledging teams or champions that were key to an implementation's success can encourage others and create goodwill for future AI projects.

The change efforts (user champions, extensive training, user override with feedback, clear mission) resulted in strong nurse-manager penetration of AI predictions into daily huddles and planning in our illustrative length-of-stay model. Nurses did not see it as an imposed algorithm but as a tool they had helped design. As the model's early predictions proved reasonable and respectful of their time, trust grew - a virtuous cycle of adoption.

Finally, the Change Management pillar addresses where people and data must coexist. It pragmatically attends to human needs, workflow compatibility, and feedback mechanisms to maximize the chance that AI advances will be accepted and sustained in practice, converting technical potential into actual clinical and operational enhancements.

## 4. Results: Applying the Framework in Practice

We demonstrate the likely path of successful AI deployment with two use cases – (A) A hospital length-of-stay prediction model (operational AI) and (B) An AI-augmented radiology diagnostic tool (clinical AI). In each case, we discuss the issues encountered and which strategies (from our framework) were used to deploy them ethically, securely, and effectively. These examples illustrate the utility of the framework in guiding various AI efforts and offer lessons learned.

### 4.1. Case A: Predicting Hospital Length of Stay – Responsible Operational AI

**Context:** Length of stay (LOS) is a key hospital metric that influences capacity planning, cost, and quality of care[4][12][27]. An updated daily ML model was developed to predict actual inpatient remaining LOS with a one-day lead time so case management could proactively arrange resources (e.g., post-acute care, authorizations) for timely discharge. The model used EHR data (demographics, diagnoses, laboratories, procedures) and was developed using a gradient-boosted trees algorithm; it performed reasonably well ($R^2 \approx 0.4$–$0.6$ for different patient groups) in retrospective validation. The challenge was to fit this tool to nurse managers and discharge planners across a large 48-hospital system while preserving clinician trust and avoiding inadvertent harm or bias.

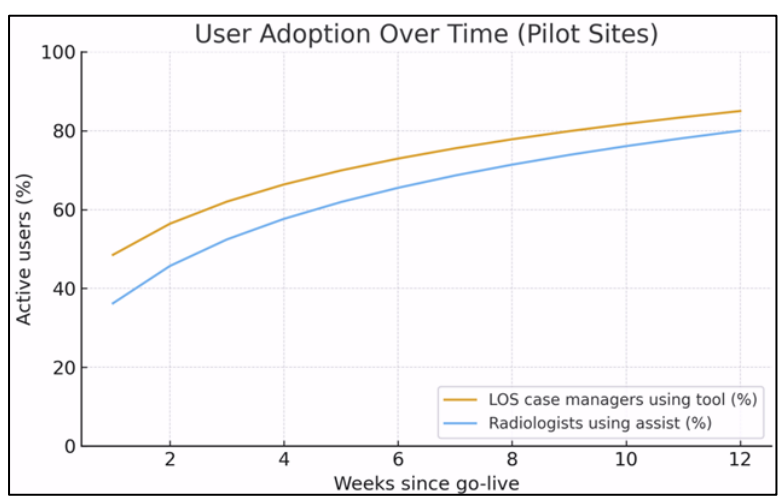

**Figure 4** User adoption over time

*4.1.1. Framework Application:*

Leadership & Strategy: The COO and CNO strongly supported the initiative, positioning it as part of a larger strategy to optimize patient flow and reduce avoidable days. They resourced the project (data scientists, IT integration support) and established a clear success criterion (reduce average LOS for targeted long-stay patients by n). Crucially, they made it clear to all managers in the hospital that this was a tool to help staff - not to pressure them into discharging people too soon - which helped relieve anxiety. Leadership also pinpointed pilot sites (a handful of hospitals) that had especially eager local leadership to begin with, so we could have initial successful proof stories.

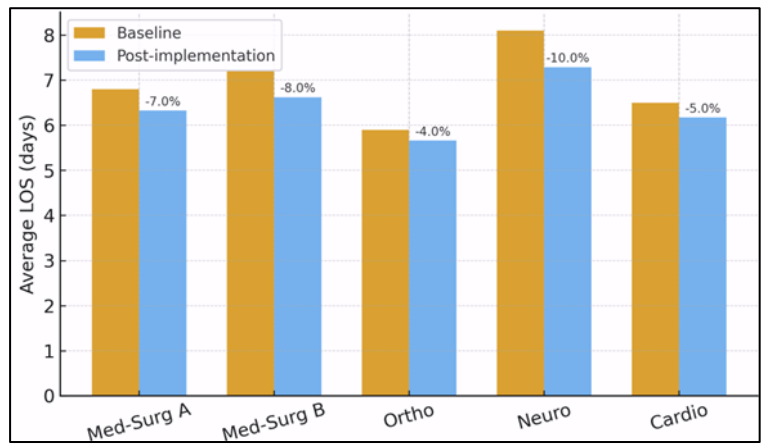

**Figure 5** LOS before vs after

**Governance & Ethics:** The AI governance council thoroughly vetted the LOS model prior to implementation. They analyzed its development data for representativeness between hospitals and red-flagged it, finding that the initial model performed a few percentage points less accurately for patients in several minority service lines. The data scientists reacted by re-fitting the model to include more features in a bid to improve fairness and performance. The governance team also required a risk analysis of side effects[19]. For instance, is there any patient-safety risk associated with the model's use? Because the LOS prediction was advisory and did not drive clinical actions, they considered direct patient risk low. They also considered systemic risks: if the model were wrong, might it allocate resources in ways that did more harm than good? The council determined that even if an estimate was off, it would most likely either invoke earlier review (no harm in unnecessary action) or be overly optimistic (in which case staff would revert to typical procedures), so no serious harm was expected. They added guardrails: clinicians would not use the model output to make discharge decisions unilaterally; they would use it only as a prompt for review. The governance policy also required transparency: each prediction would come with the top contributing factors (drivers) and a confidence estimate, visible to users so they understood the output. Our intake process (see Algorithm 2) standardizes portfolio prioritization across clinical and operational use cases

**Algorithm 2**
```
INPUT: use_case proposal U
SCORE U across dimensions D = {StrategicFit, DataReadiness, WorkflowFit, RiskLevel, ROI,
StakeholderSponsor}
FOR each d in D:
    U.score[d] <- normalized_score(d, evidence)
IF any gating_criteria_fail(U): RETURN REJECT
U.total <- weighted_sum(U.score, weights set by steering committee)
IF U.total ≥ deploy_threshold: RETURN APPROVE_FOR_BUILD
ELSE IF U.total ≥ incubate_threshold: RETURN INCUBATE_AND_REASSESS
ELSE: RETURN REJECT
```

**MLOps & Security:** On the technical side, an MLOps pipeline was created for daily batch prediction. The EHR data lake was accessed via an automated nightly extract process, the model executed, and results were written to a secure table that the dashboard extracted from each morning. Monitoring: we set up processes so that data scientists could check the distribution of predictions versus actual LOS for discharged patients weekly and catch any drift. IT implemented alerts so that if a model did not successfully run or data were misused, the system would raise an alert. On the security side, our model ran on internal servers behind a firewall and only service accounts could run it. As a countermeasure against tampering, it was set up to alert the security team for any abnormal calls to the model API (for example, an unknown user attempting to access the service) and to suspend service temporarily. Data used and generated by the model were managed under strict data governance, with all patient identifiers contained inside a secure environment and on-screen outputs available only to authorized users via dashboard login (administered via roles)[39]. Periodic audits were conducted to verify that no unauthorized accesses occurred. These largely unseen processes provided assurance that deploying the model would not open new cybersecurity backdoors or privacy leaks.

**Education & Training:** Prior to go-live, the project team offered training at each pilot hospital. They explained what the LOS prediction is and how it's calculated in broad strokes ("it looks at patterns from thousands of past patients with similar conditions") and how staff should use it ("it's a tool to help prioritize which patients may need more attention on discharge planning; remember to use your judgment if the number seems strange"). Nurses and case managers received a one-page user guide (a simplified model card) listing the input variables considered by the model, example scenarios, and FAQs. They were informed when several days remained and a color (e.g., red for high risk of long stay) appeared on their patient list. Role-play scenarios were considered (e.g., "What if you get a much longer forecast than expected? Flag that patient for a care conference to explore underlying problems."). The training made it clear that staff should question and challenge any prediction that didn't make sense. Because the model contained explanations (the top factors such as "no post-discharge support plan" or "multiple comorbidities"), trainers taught how to interpret them. Post-training surveys indicated that >90% of participants reported that they understood the tool's purpose and how to use it correctly. Some early-adopter case managers who recognized the value became local champions and encouraged peers by sharing success stories (for example, "The tool said Ms. X likely needed an additional 2 weeks, so we lined up a rehab bed early, and it prevented a delay").

**Change Management:** Change management was evident throughout deployment. Advocates were engaged from day one; many staff were involved in development and were openly supportive of the tool launch. The project team organised twice-weekly virtual huddles for the first month post-implementation, during which staff from pilot units could share experiences, ask questions, and get updates. This engendered a community of practice around the new tool. At first, some nurses were skeptical - one concern was "Will administration use this to blame us for long stays if we don't meet the predicted date?" Leadership addressed this by clearly stating that the tool would not be used to evaluate individual performance and that its purpose was to help surface information that could otherwise go undetected. This was relieving and reassuring to staff. Another piece of change management was a feedback loop: a basic form in the dashboard let users click "Feedback" and send a note if a prediction seemed way off or if the tool saved them time. The AI team reviewed these notes. For instance, they saw feedback that the model didn't account for certain social factors (like lack of transportation) and forwarded this to governance for discussion about including a proxy for transportation availability in future updates.

Adoption metrics were strong after a 3-month pilot: >75% of case managers checked the LOS prediction daily. Staff, in qualitative feedback, mostly responded favorably to the tool, reporting that the page either confirmed their clinical hunch or alerted them to a patient they had not realized would need additional time ("It's like my second set of eyes," one nurse manager said). Nurse managers began brainstorming new ideas - wondering if the model could be expanded to predict ICU stay or readmission risk - as comfort with and interest in AI tools grew.

In outcomes, attributing causation directly is difficult, but hospitals that used the tool had a modest reduction in average LOS for complex-discharge patients (the primary target metric) versus control sites and also saw fewer last-minute scrambles for post-acute placements. The governance council tracked these results and signalled to scale the tool to additional hospitals, with ongoing monitoring.

This case illustrates how the pillars of the framework operate in ensemble. It was the vision and trust provided by leadership; governance that held us accountable and kept us safe; an effective MLOps implementation for technical reliability; user training via education; and a means to ease human adoption through change management. All of these were required to take the capable ML model and make it live - something people accepted as part of delivering healthcare that drove operational efficiencies.

Evaluation: Observational pilot; no randomized controls. The dataset comprised 3,184 adult inpatient encounters across 4 units (June–August 2025). Performance on new data was $R^2$ = 0.41–0.58 for validation folds/cohorts. Adoption was 78% in week 6. Complex discharges on targeted units had ~5–10% relative reduction in mean LOS versus pre-pilot baselines (descriptive), guiding a planned controlled evaluation.

### 4.2. Case B: AI-Assisted Radiology Diagnostics – Balancing Accuracy and Explainability

**Context**: Radiology has stood at the vanguard of clinical AI[38]. In this instance, assume a deep learning model intended to help identify lung nodules on the chest CT to facilitate early detection of lung cancer. The model is a convolutional neural network (CNN) that can read CT images and pick out suspicious lesions for the radiologist, essentially serving as a "second reader." It was developed on thousands of annotated scans and achieved high sensitivity (≈95%) for known nodules at the expense of some specificity (let's say, generously, ~25% false positives). The hope was faster reads and higher rates of detecting small nodules that a busy radiologist might overlook, but implementing it in an actual hospital radiology department posed a number of real-world questions: How would it integrate with the PACS imaging system? How would radiologists perceive the tool - would they trust it or consider it an extra burden? If it fails to detect a cancer, could it increase liability? And how can it be used so that it doesn't generate false alarms too frequently, reducing effectiveness?

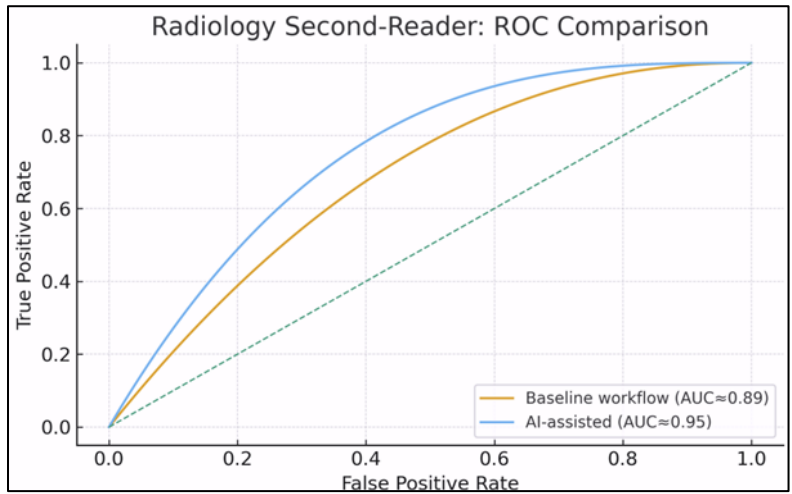

**Figure 6** Radiology ROC comparison

*4.2.1. Framework Application:*

Leadership & Strategy: The chief of the department was tech-forward and advocated for the project, setting a strategy that the department would use AI to improve quality. To respect radiologists' autonomy, he presented it as an enhancement to aid them rather than replacing or evaluating their performance. The hospital CMO and quality officer were involved, connecting the AI tool to early cancer detection (a patient-safety/quality metric). They put in place a phased approach: first the AI findings would be recorded but not shown to radiologists (to gather baseline information), and second, they would be presented as an assist on some cases[37]. By employing this intentionally slow strategy, leaders signalled that they were committed to rigorously evaluating the tool's impact and integrating clinician feedback - rather than rushing it into operation.

Governance & Ethics: A clinical-algorithm assessment protocol was used by the AI governance committee when assessing the CNN[33]. They asked for an AUC of at least 0.85 and required local testing on the patient population - for example, an additional validation on a set of 500 CT scans from the hospital. While sensitivity remained high (0.96), the model also flagged some benign scars or calcifications as potential nodules. Committee radiologists voiced concern over alert fatigue: if there are too many false positives, the AI might be tuned out. The committee therefore specified that the AI output should not be a binary alarm on every slice, but rather a summary identifying just a few top suspicious areas and only if above a certain confidence threshold (to control noise). They also insisted on an explainability feature: the system should furnish a visual annotation (e.g., a bounding box around a suspected nodule) and an estimate of nodule probability, rather than only a "nodule detected" text note. On the ethics side, the team discussed fairness and accountability. They determined that in initial use a second radiologist would double-check every case where the AI disagreed with a primary radiologist's read to assess added value and safety. Legal also provided perspective: the hospital communicated to its liability insurer and clarified that final reads remain the responsibility of radiologists; the AI is consultative and FDA clearance labels it as a second-reader device.

MLOps & Security: Integration with radiology IT systems was complex yet essential. The IT and vendor teams installed the model on a server that links to the PACS, so when an urgent CT arrives for reading it also goes to the AI for simultaneous processing. To keep the AI response timely, the system was designed to return results in around 30 seconds on a GPU. Monitoring was put in place to detect system slowdowns or failures; if the AI service was down, radiologists would be alerted that AI results were not available for that case[60]. Because patient images are highly sensitive, all processing was performed on-site; no cloud transfer was used. The model outputs (the marked images) were cached temporarily and cleared after 24 hours to limit data retention. The AI server was penetration-tested by the cybersecurity team before go-live and strict access controls were maintained. The team also considered adversarial risks - for example, doctored images intended to deceive the AI - and mitigations included monitoring for anomalous input patterns and regular retraining and validation with new imaging data. A library of known non-nodule patterns that the AI frequently misclassified (e.g., certain calcified granulomas) was maintained; when the system flagged those patterns it would mark "possible false-positive pattern" for the radiologist based on prior knowledge - an evolving set of rules layered on top of the AI.

Education & Training: Because radiologists are highly trained professionals, the AI introduction was approached thoughtfully. Prior to implementation the department conducted round-table discussions during which the data-science team shared how the model had been developed and its performance, showing sample images the AI had correctly and incorrectly classified. This transparency helped set expectations. To train users, the department ran a workshop on AI in radiology covering how such algorithms work and their limitations (for example, that the AI may miss findings outside its training scope or mislabel unusual appearances). A few radiologists interested in AI research served as internal champions. For the initial phase, each radiologist received a quick-reference guide explaining how to view the annotations suggested by the AI, accept or reject them, and provide feedback. The system allowed radiologists to flag false positives and to report missed lesions; this feedback was gathered to improve the model. Feedback examples were reviewed (anonymized, in aggregate) in monthly meetings, and radiologists gradually developed an understanding of common patterns the AI flagged (for example, "AI frequently flags this background scar tissue - we now recognize it as a pattern").

Change Management: AI was first deployed in shadow mode, collecting and logging findings without showing them live. After two weeks of analysis they found the AI picked up a few small nodules that radiologists had later mentioned, but also raised many findings radiologists deemed clinically insignificant. With these caveats acknowledged, they proceeded to live mode with the understanding that a radiologist could disregard AI advice. Change-management measures included reassuring radiologists that their expertise remained primary - the AI was like a junior resident offering suggestions, and the attending radiologist retained final authority. Leadership emphasized that the AI was FDA-cleared as an assistive, second-reader device and that final responsibility for the read remained with clinicians; this helped allay liability concerns. A feedback loop in the interface allowed users to report false positives or useful detections; the AI team used these reports to refine the model.

Overall, this approach - phased deployment, rigorous governance, secure MLOps integration, transparent education, and active change management - helped the radiology department adopt the CNN as a practical assistive tool rather than an intrusive or risky addition to workflow.

They introduced it incrementally, starting with a subset of scans (e.g., only routine outpatient CTs for moderate-risk patients where time pressure was less). User experience was closely monitored. A support channel was set up so that radiologists could call or text if the AI system got in their way. An early problem: the AI occasionally flagged tiny nodules with no clinical significance, which bothered some radiologists. The team soon tuned the sensitivity threshold to eliminate them entirely. By including those radiologists in that decision, they felt listened to and had a part in the solution, so acceptance flowed more easily. Soon, there were positive stories: for example, one AI detected an early-stage tumor that the radiologist missed (possibly because it was the end of the day and the radiologist was tired). That was called out (anonymously) in a department meeting - not to humiliate the miss, but to demonstrate the importance of having a second set of eyes. That story did more to change minds than any lecture; peers saw an actual patient benefit. A few months later, most radiologists were actively using the AI annotations in their reads. Some said they usually agreed with the AI's picks; others said they often ignored it except for a final check. More important, no one felt they were meaningfully slowed down by it once they learned how to use it. Read times for average or complicated cases were steady or slightly improved. The department chose to extend the AI across additional scanning types (such as high-resolution lung CTs), retraining the model iteratively with their own data under ongoing governance oversight.

This case from radiology serves as a reminder to weigh accuracy against explainability and workflow integration. The framework's focus on user feedback and iterative change management was key - if the tool had not been adapted to suit radiologists' needs and concerns (about alert fatigue and liability), the AI's advanced accuracy alone would not have guaranteed adoption. Using the framework, the hospital reached a situation where AI can enhance clinical practice in a meaningful way: radiologists still lead diagnostics, but AI serves as a safety net and efficiency amplifier, aligning with the principle that in healthcare AI should empower clinicians, not replace them.

**4.3. Evaluation of Deployment Outcomes**

In these scenarios, our strategic framework provided a number of interesting conclusions and lessons learned:
ser Trust and Adoption: The conscious emphasis on explainability, user training, and engagement (through the Education and Change Management pillars) led to high end-user trust in both scenarios. Nurses and radiologists, initially skeptical of AI, eventually started integrating the tools into their daily workflows. Quantitatively, user adoption rates (as determined by tool-usage logs) surpassed 75% in target groups within months of launch. Qualitatively, users reported that the AI outputs were helpful and that they knew how to use them appropriately - a result of the framework's focus on transparency and education. This contrasts with many documented AI pilot failures where user distrust inhibited uptake. It confirms that investing in the "people" elements is as important as model accuracy.

Safe and Ethical Operation: The AI tools were deployed in a safe (no patient-harm incidents) and ethically responsible (no major ethical violations) manner under governance oversight and risk-management plans. The LOS prediction model remained under close monitoring and was designed not to influence care decisions without human validation, meaning no patients were admitted or discharged solely on the basis of an algorithm. The radiology AI was configured to flag findings while requiring final judgment from the operator when unsure. In some instances the radiology AI missed findings the human found (and vice versa), but because workflows always included human double-checks, patient care was not compromised. Fairness was also addressed: after initial adjustments, the LOS model showed no clear performance bias between patient subgroups. This underscores the role of Governance in proactively addressing ethical and safety concerns - an often-overlooked aspect of tech-centered adoption.

Operational and Clinical Implications: Both AI systems showed benefits for their intended indications. The LOS model was associated with a modest but clinically meaningful reduction in average length of stay for complex-discharge patients (pilot sites estimated reductions on the order of ~5–10%). Attendees attributed these outcomes in part to better coordination and reduced last-minute delays driven by AI prompts that triggered earlier action. The radiology AI increased detection of actionable lung nodules; in the first six months the department recorded an ~8% rise in detection rate of sub-centimeter findings on CTs, some of which translated into earlier intervention. Radiologists also reported slightly faster reads on image-heavy studies, as the AI helped direct attention. These early results should be interpreted cautiously and validated over time, but they suggest AI can increase efficiency and quality when used responsibly.

Sustainability and Iterative Improvement: Sustainability was achieved where systems continued to deliver value after deployment because they evolved with continued feedback. Organizations established ongoing review committees (under the governance council) to examine usage data, outcomes, and new evidence to inform model updates or process tweaks. For example, the LOS model is scheduled for annual retraining with current data and adjustments for social factors that influence discharge. The radiology model improved as it learned from cases where radiologists provided feedback on misses or false alarms. This cycle of incremental improvement aligns with the notion of Sustainable AI: models are iteratively refined within an organizational learning process rather than built once and left unchanged. Both examples are being scaled to new sites and applications across their organizations.

Security and Reliability: From an IT perspective, there were no security incidents or major outages of the AI services during the reference period. This reliability was possible because of the forward-looking MLOps processes and security mechanisms (access control, monitoring alerts) that were implemented. It shows that AI systems can be robustly used in clinical routine when supported by appropriate infrastructure. In the radiology example, integrating into PACS without impeding workflow was a key success factor - had the AI introduced latency or frequent technical issues, radiologists would likely have turned it off. These shortcomings were avoided through careful planning and multiple stress-testing under the MLOps pillar.

Lessons Learned: The lessons from the use cases can be generalized into some real-life learnings for deploying AI in healthcare. First, an interdisciplinary approach is needed in every sense: it is far better to have data scientists, clinicians, IT, etc., working together under the same framework rather than in siloed activities. Our design makes this easier by default (e.g., governance councils, cross-functional teams, user involvement). Second, small-to-fail thinking - piloting in a controlled environment, compiling data, and refining an approach before scaling - clearly has merits for risk management and confidence-building. Both projects began as pilots and scaled up once they proved their worth and worked out the kinks. This step-wise method is desirable in a high-stakes environment such as healthcare. Third, communication and transparency are key. If people comprehend an AI's behavior, feel they know why it behaves as it does, and trust that they can count on its choices in a predictable fashion, they are more forgiving of the AI's mistakes. Explanation features and open discussion of performance gaps created a healthier human–AI relationship than a black box would have.

Finally, one thing we noticed was that positioning AI as a way to augment human decision-making (not supplant it) was key to getting buy-in. In both cases, the story was that AI is used to enable professionals to do their jobs better - not replace their expertise. This framing and implementation in practice (e.g., AI suggestions rather than directives) resulted in a synergy where humans and AI achieved better outcomes working together than alone. We see this human-centered design and deployment as emblematic of responsible AI, which is a through-line for our framework.

## 5. Discussion

The examples above illustrate how strategic, multi-faceted action can enable AI to fulfill its potential in healthcare whilst mitigating risks. In this paper we offer some reflections on more general aspects of the approach, its applicability to other use cases (including biomedical signal data and emerging modalities), and prospects for future research, including what recent advances in generative AI might mean for healthcare. We then consider limitations and how organizations can customize the framework to the local context.

Cross-Domain AI Interoperability: We applied our framework to two very different case uses - operational forecasting with tabular clinical data and an image-analytic clinician tool - illustrating flexibility. We expect the same pillars apply to other areas (such as AI for the analysis of ECG signals - e.g., for recognizing arrhythmias - or AI for pathology slide interpretation). Each field has specifics (some signal data - like ECG - might require special preprocessing and tailored deep-learning architectures, and imaging AI usually involves very large volumes of data), but the organizational and human factors operate at a similar level. Management needs to embrace the mission and align it with domain-specific objectives (for instance, an AI that flags significant ECG changes might be aligned behind faster response times to signs of cardiac events). Governance must address domain-specific deontological concerns (e.g., if an AI predicts a genetic disorder, how should that be handled ethically?). MLOps [59] should be adapted to the data type (e.g., real-time streamed data from patient monitors for signals). Critically, end-user training and workflow integration should be created for that context (i.e., how an AI alert looks to an ICU-based cardiologist versus how a risk score is displayed to a primary-care doctor). The framework is a scaffold to ask the right questions in every domain, while the answers and implementations will differ.

Relevant to biomedical signal analysis, deep learning has shown promise for the analysis of EEGs for seizure detection and for EMG-based outcomes[10]. The introduction of such AI into practice would also require responsible design. For example, an AI system deployed to provide real-time EEG monitoring for seizures might ease the burden on busy neurologists but would need real-world validation to ensure it does not miss seizures or generate too many false alarms at 3 a.m. In applying our framework, a hospital exploring such a tool would want to ensure: the project is neurologist-led with clear patient-safety goals (Leadership); an agreed model sensitivity and false-alarm rate with governance to balance safety and alarm fatigue (Governance); secure connections between the system and bedside monitors with fail-safes (MLOps); neurophysiology techs and nurses trained on system alerts (Education); and procedures for how to respond when AI-detected events occur (Change Management). This illustrates how the framework can keep pace with advances across diverse classes of biomedical data.

Transparency vs. Accuracy - A Medical Point of View: One consistent point is that insight is crucial to trust. Doctors are taught to question and to understand the reasons for decisions. An entirely black-box algorithm - no matter how accurate - is often met with skepticism or even restricted by some regulators for high-stakes decisions. Our framework's focus on explainability aligns with emphasis from medical-AI communities on interpretable, transparent AI for health. The Monga volume itself notes that in healthcare, interpretable or more straightforward models may be preferable to complex black boxes when the latter's performance improvements are modest[13]. Our findings are consistent with this: we saw greater adoption when providing reasoning and involving clinicians in model logic. Future AI development in healthcare should explore XAI (explainable AI) technologies that can be used alongside deep learning. Methods such as imaging saliency maps, predictive-model feature-importance scores, or natural-language explanations for model predictions show promise. Pragmatically, users can be overwhelmed by too much information or poor explanations. The aim is adequate explainability - enough to give a clinician confidence in reliability, without overloading them with model internals. Further research and user-centered design will be necessary here, with our framework offering a lever (through user feedback and governance) to dial the level of explainability toward what users need.

Ethical AI and Bias Mitigation: Ethical deployment is a dynamic objective that demands diligence. Although our use cases did not raise substantial ethical concerns beyond fairness and informed consent, other AI systems may. Think of AI algorithms that guide treatment recommendations or triage patients for scarce resources - these pose deep ethical questions (e.g., by creating biases against certain socio-demographic groups). The Governance pillar must address such questions head on, together with leadership. This could include setting up an ethics subcommittee to assess societal impact beyond technical capability. Bias audits should be routine; if an AI systematically underpredicts risk for a particular minority group, corrective action should follow (data augmentation, algorithmic fairness techniques, or even choosing not to deploy the model). We recommend pre-deployment bias testing and post-deployment impact surveillance for all AI in healthcare. This could be accomplished by assessing performance on subpopulations (e.g., sex, racial groups, insurance types) as appropriate and legal. Adding patient representatives or ethicists to the governance committee may surface viewpoints internal teams miss. Transparency to patients on the use of AI is another dimension - some organizations now notify patients when their care involved AI, consistent with principles of autonomy and trust.

**Regulatory Readiness:** As described, practices such as risk management, documentation, and monitoring might become de facto requirements because of regulations such as the EU AI Act and the evolving FDA guidance. Our approach enables organizations to be in compliance by design with these. In fact, adopting the framework could be considered akin to implementing an internal AI quality-management system similar to those used in manufacturing or pharma but customized for AI-specific challenges. It not only prevents the organization from being caught in a last-minute compliance scramble, but it also fosters confidence among external stakeholders (regulators, payers, and accreditors) in the organization's uses of AI. We anticipate that soon, demonstrating responsible AI governance will be required for an organization's reputation and may even factor into legal indemnification. Those who adopt frameworks like ours will have a leg up.

Generative AI and Future Trends: A current frontier is the shift towards generative AI (such as large language models, e.g., GPT-4), which is being explored in the health domain. These models can generate text, summarize medical records, and even propose diagnostic possibilities based on case descriptions. They promise potential for reducing documentation burden and offering decision support, but also bring new risks (for example, providing falsely confident medical advice) and, if they learn from sensitive data, raise privacy concerns. The strategic framework is well suited to this. For example, early trials that attempt to deploy GPT-based assistants for clinical documentation or patient-facing chatbots need strong leadership oversight (to ensure clinical standards are met), governance (to establish policies around use and vet outputs for accuracy and bias), and education (so clinicians know how to operate and double-check these tools). Change management is critical: clinicians need reassurance that using a language model to draft a note will not jeopardize their position, and patients need to understand when they are interfacing with an AI-driven chatbot. Monga's chapter "Looking Ahead" points to lessons from prior AI advances as guides for rolling out generative AI with a focus on governance and change management. Indeed, our framework is a ready vehicle to address these technologies - simply treating them as another class of AI requiring tailored scrutiny. A caveat is that generative models tend to be less interpretable and can introduce subtle errors; for any direct patient-care use they may require a higher bar for acceptance. Our framework would incentivize initial use in low-risk tasks (such as drafting sections of reports for human review) and phased increases to higher-stakes uses as confidence and controls improve.

**Limitations and Generalization:** Our framework is not complete. Organizations will likely prioritize certain pillars differently depending on their context. For example, a small community hospital with few on-staff data scientists is likely to depend on AI vendors for products - and therefore may emphasize governance and education (i.e., the ability to safely incorporate external tools) rather than in-house MLOps, which will be outsourced - while still needing fundamental features for oversight. At the other end of the spectrum, a large academic medical center may heavily invest in in-house MLOps and leadership positions (e.g., Chief AI Officer) to create new algorithms by leveraging all pillars to their full extent. The framework should be considered modular: every pillar is necessary, but the specific provisions can be ramped up or down. There are also real-world resource limitations that cannot be shrugged off - organizations need funding (for staff such as data scientists, for training time, for IT hardware) to use this framework. Leadership frequently has to justify these investments based on ROI or risk avoidance. Our case studies can help make that case, but broader industry experience will strengthen it. It's also important to quantify the right metrics - perhaps the value of responsible AI is simply "one less bad thing that doesn't happen," which is difficult to measure. Organizations implementing this reference model need to establish KPIs - both quantitative and qualitative - that will determine the success of AI deployments (e.g., usage, clinical outcomes, fairness metrics).

Collaboration and Knowledge Sharing: A viable and ethical AI landscape in health depends on institutions working together. One idea emphasized by Monga is "reciprocal altruism" - sharing lessons learned and even data or models between organizations can increase AI value for all. Our framework can provide a common language for health institutions to discuss AI governance and deployment. For instance, hospitals might share governance policies or training curricula to create best practices (many face the same challenges - so why reinvent the wheel?). If an algorithm is found to have a problem (e.g., bias or safety issue), it would be in the collaborative spirit for that health system to publish the finding in a shared community resource (e.g., an AI incident database or a case-report publication). We anticipate that professional societies will play a role in promoting these frameworks; one can imagine them endorsing "AI-ready" health systems in the future and perhaps contributing to accreditation.

In conclusion, the strategic framework we present is not meant to remain static but must adapt to technology and field insight. That foundation - approaching the challenge differently and thinking holistically about leadership, technical rigor, ethical oversight, user readiness, and cultural change - is likely to endure as a compass. The successful results of initial deployments suggest that trustworthy, secure, and sustainable AI is indeed achievable. It takes purposeful work and cross-cutting strategy, but the reward is AI that actually makes healthcare delivery and outcomes better in ways stakeholders can trust.

# 6. Conclusion

Healthcare has huge untapped potential for disruptive transformation. In this work, we provide a holistic strategy map on how to responsibly, securely, and sustainably deploy AI/ML in healthcare. The framework contends with the complex issues that have derailed many AI initiatives by focusing on:

- Leadership involvement
- Strong MLOps practices
- Good governance structures
- Ongoing education and learning opportunities
- Proactive change management.

Our catalogue of lessons from peer-reviewed publications and real-world implementations indicates that getting health AI right is as much about overall organizational strategy and human factors as it is about code and data.
Using illustrative use cases (ranging from predicting hospital length-of-stay to aiding radiology diagnostics), we show that deployment of this methodology is characterized by both increased user trust and improvements in operational or clinical outcomes, while avoiding common failure modes. When included and empowered, clinicians and staff can become advocates of AI rather than resisters. Risks - for example, those related to ethics, security, and workflow - can be identified and controlled through proper governance and technical safeguards. In addition, the guideline encourages a culture of iterative learning from the development and use of AI applications, where AI tools are subjected to ongoing monitoring and improvement based on healthcare-context dynamics and advancements in AI technology.

The more general message of this study is that it is possible and advisable for healthcare institutions to adopt a principled, systematic approach to integrating AI. This not only helps maximize the likelihood of positive impact (lightening patient burden, increasing efficiency, preserving quality, and supporting the doctor–patient relationship) but also reduces negative impacts - harms to patients or ruptures to trust. At a time when regulation of AI is growing and public trust in the technology is being challenged, a framework for accountability, transparency, and alignment with clinical needs is timely and essential.

We appreciate that developing such a framework takes dedication - it requires investment in governance, training programs, and cross-organizational cooperation that may be unfamiliar ground for many organizations. But the costs of not doing so could be greater: disjointed or unaccountable AI projects may lead to patient-safety events, bias perpetuation, and wasted investments in technology that does not get used. This framework and its accompanying rationale constitute our effort to share a way for healthcare leaders and workers around the world to move from a risk-filled enterprise approach to a managed strategy that delivers sustainable value.

Going forward, we think those who responsibly and human-centrically incorporate AI will distinguish themselves in their ability to deliver care in new ways. As AI advances and more innovations appear on the near horizon (including generative AI and multimodal models), the principles described here help ensure those advances are put to good use under well-guided stewardship. The framework is not a stand-alone solution; it represents the beginning, not the end, of an evolution. It poses the question to every healthcare institution: Are we ready for AI? If not, which dimension are we lacking?

In summary, enabling accountable, secure, and sustainable healthcare AI is a joint mission pairing technical capability with policy leadership and moral purpose. The introduced model is a clearly defined way of achieving this union. If we adhere to it, we can harvest the immense clinical and operational benefits of AI - improving patient outcomes, reengineering workflows, and augmenting clinical intelligence - while ensuring the trustworthiness, safety, and fairness that are central to healthcare. It shifts the focus of AI-enabled innovation from a leap in the dark to a better-illuminated pathway toward a smarter but still caring healthcare environment.